\documentclass[amsmath,amssymb]{revtex4}

\usepackage{graphicx}

\begin{document}

\title{Secondary-Structure Design of Proteins by a Backbone Torsion Energy}


\author{%
Yoshitake Sakae$^{1,}$\footnote{e-mail: sakae@hiroshima-u.ac.jp} and
Yuko Okamoto$^{2,}$\footnote{e-mail: okamoto@phys.nagoya-u.ac.jp} 
}


\affiliation{%
$^{1}$ Faculty of Science \\
Hiroshima University \\
Higashi-Hiroshima, Hiroshima 739-8530, Japan \\
$^{2}$ Department of Physics \\ 
Nagoya University \\
Nagoya, Aichi 464-8602, Japan\\
}

\begin{abstract}
We propose a new backbone-torsion-energy term in the force field for protein systems. 
This torsion-energy term is represented by a double Fourier series in two variables,
the backbone dihedral angles $\phi$ and $\psi$.
It gives a natural representation of the torsion energy in the Ramachandran space
in the sense that any two-dimensional energy surface periodic in both $\phi$ and $\psi$ can be
expanded by the double Fourier series.
We can then easily control secondary-structure-forming tendencies by modifying the torsion-energy surface.
For instance, we can increase/decrease the $\alpha$-helix-forming-tendencies by lowering/raising the torsion-energy surface 
in the $\alpha$-helix region and likewise increase/decrease the $\beta$-sheet-forming tendencies by lowering/raising the surface 
in the $\beta$-sheet region in the Ramachandran space.
We applied our approach to AMBER parm94 and AMBER parm96 force fields and demonstrated
that our 
modifications of the torsion-energy terms resulted in the expected changes of 
secondary-structure-forming-tendencies by performing folding simulations of 
$\alpha$-helical and $\beta$-hairpin peptides.
\end{abstract}


\maketitle
   
\section{Introduction}
\label{introduction}
A force field (or potential energy) for protein systems is necessary to perform molecular simulations 
based on Monte Carlo (MC) and molecular dynamics (MD) methods. 
Well-known force fields are, for instance, 
AMBER \cite{parm94, parm96, parm99}, CHARMM \cite{charmm}, OPLS \cite{opls1,opls2}, 
GROMOS \cite{gromos}, and ECEPP \cite{ECEPP}.
These force fields have been parameterized to fit experimental
data of small molecules and, for some terms, the results of quantum chemistry calculations
and MD simulations.
We have recently carried out detailed comparisons of three versions of AMBER
(parm94 \cite{parm94}, parm96 \cite{parm96},
and parm99 \cite{parm99}), CHARMM \cite{charmm},
OPLS-AA/L \cite{opls2}, and GROMOS \cite{gromos}
by generalized-ensemble simulations \cite{GEA} of two small peptides in explicit solvent. \cite{YSO1,YSO2}
The results of these comparisons indicated that these force fields yield quite different 
secondary-structure-forming tendencies.
It was shown that AMBER parm94 is the most (and too much) $\alpha$-helix-forming among
the six force fields studied and that AMBER parm99 and CHARMM give ample amount of
$\alpha$-helix structures, whereas AMBER parm96, OPLS-AA/L, and GROMOS are more
$\beta$-sheet-forming than the rest. \cite{YSO1,YSO2} 
These results were confirmed by the folding
simulations of the two peptides with implicit solvent model. \cite{SO1,SO2,SO3}
Our conclusion was that we need to refine and improve the existing force-field
parameters in order to yield the secondary-structure-forming tendencies that 
agree with experimental implications. \cite{YSO1,YSO2,SO1,SO2,SO3}
Among the force-field terms, the torsion-energy term is the most 
problematic.  For instance, the parm94, parm96, and parm99 versions of AMBER differ mainly in 
the backbone-torsion-energy parameters.
The main changes from OPLS-AA to OPLS-AA/L can also be found in the torsion-energy term.
\cite{opls2}
Other trials of force-field refinement mainly concentrate on the torsion-energy terms.
These modifications of the torsion energy are usually based on 
quantum chemistry calculatioins \cite{Carlos,Duan,IWA,MFB,Kamiya}.
It was also proposed to set the backbone torsion-energy term simply to zero. \cite{GS}
We have proposed another force-field refinement procedures \cite{SO1,SO2,SO3}.
Using the standard functional forms of the force fields, we refine the
force-field parameters so that they are the most consistent with the coordinates of
proteins in the Protein Data Bank (PDB) database.  This is achieved by minimization of forces acting
on the atoms with the coordinates of the PDB with respect to the force-field parameters.
We found that our refinement of partial-charge and backbone-torsion-energy
parameters resulted in the improvement of
three versions of AMBER (parm94, parm96, and parm99), CHARMM, and
OPLS-AA \cite{SO1,SO2,SO3}.

In this article, we propose a new backbone-torsion-energy term, which is 
represented by a double Fourier series in two variables,
the backbone dihedral angles $\phi$ and $\psi$.
This expression gives a natural representation of the torsion energy in the Ramachandran space 
\cite{Rama_Sasi}
in the sense that any two-dimensional energy surface periodic in both $\phi$ and $\psi$ can be
expanded by the double Fourier series.
We can then easily control secondary-structure-forming tendencies by modifying the backbone-torsion-energy surface.
We accommodated our approach to AMBER parm94 and AMBER parm96 and tested whether our modifications of the backbone-torsion-energy term resulted in the expected change of secondary-structure-forming-tendencies by performing folding simulations of $\alpha$-helical and $\beta$-hairpin peptides.



In section 2 the details of the double Fourier series as the backbone-torsion-energy term 
are given.
In section 3 the results of applications of the double Fourier series to
modifications of AMBER parm94 and parm96 force fields are presented.
Section 4 is devoted to conclusions.

\section{Methods}
\label{method}
The existing all-atom force fields for protein systems such as AMBER
and CHARMM use essentially the same functional forms for the conformational
potential energy $E_{\rm conf}$ except for
minor differences.  $E_{\rm conf}$ can be written as, for instance,
\begin{equation}
E_{\rm conf} = E_{\rm BL} + E_{\rm BA} + E_{\rm torsion} + E_{\rm nonbond}~.
\label{ene_conf}
\end{equation}
Here, $E_{\rm BL}$, $E_{\rm BA}$, $E_{\rm torsion}$, and $E_{\rm nonbond}$
represent the bond-stretching term, the 
bond-bending term, the torsion-energy term, and the nonbonded energy term, respectively.
The torsion energy is usually given by 
\begin{equation}
E_{\rm torsion} = \sum_{{\rm dihedral~angle}~\Phi} \sum_n \frac{V_n}{2} [ 1 + \cos (n \Phi - \gamma_n) ]~,
\label{ene_torsion}
\end{equation}
where the first summation is taken over all dihedral angles $\Phi$ (both in 
the backbone and in the side chains), 
$n$ is the number of waves, $\gamma_n$ is the phase, and $V_n$ is 
the Fourier coefficient.
Separating the contributions $E(\phi,\psi)$ of the backbone dihedral angles $\phi$ and $\psi$ 
from the rest of the torsion terms $E_{\rm rest}$, we can write
\begin{equation}
E_{\rm torsion} = E(\phi,\psi) + E_{\rm rest}~,
\label{ene_torsionb}
\end{equation}
where we have
\begin{equation}
E(\phi,\psi) = \sum_m \frac{V_m}{2} [ 1 + \cos (m \phi - \gamma_m) ]
+\sum_n \frac{V_n}{2} [ 1 + \cos (n \psi - \gamma_n) ]~.
\label{ene_torsion2}
\end{equation}
For example, the coefficients for the cases of AMBER parm94 and AMBER parm96
are summarized in Table~\ref{table-org-torsion}, and we can rewrite
$E(\phi,\psi)$ in Eq.~(\ref{ene_torsion2}) as follows:
\begin{equation}
E_{\rm parm94}(\phi,\psi) 
= 2.7 - 0.2 \cos 2 \phi - 0.75 \cos \psi - 1.35 \cos 2 \psi - 0.4 \cos 4 \psi~,
\label{ene_torsion_parm94}
\end{equation}
\begin{equation}
E_{\rm parm96}(\phi,\psi) 
= 2.3 + 0.85 \cos \phi - 0.3 \cos 2 \phi + 0.85 \cos \psi - 0.3 \cos 2 \psi~.
\label{ene_torsion_parm96}
\end{equation}

The backbone-torsion-energy term $E(\phi,\psi)$ in Eq.~(\ref{ene_torsion2})
is a sum of two one-dimensional Fourier series: 
one is for $\phi$ and the other for $\psi$.
The two variables $\phi$ and $\psi$ are independent, and
no correlation of $\phi$ and $\psi$ can be incorporated.
Any periodic function of $\phi$ and $\psi$ with period $2\pi$ can be expanded by 
a double Fourier series.
As a simple generalization of $E(\phi,\psi)$, we therefore propose to express 
this backbone torsion energy by
the following double Fourier series:
\begin{eqnarray}
{\cal E}(\phi,\psi) ~~=~~ a & + & \sum_{m=1}^{\infty}(b_{m} \cos m\phi + c_{m} \sin m\phi) \nonumber\\ 
& + & \sum_{n=1}^{\infty}(d_{n} \cos n\psi + e_{n} \sin n\psi) \nonumber\\ 
& + & \sum_{m=1}^{\infty} \sum_{n=1}^{\infty} (f_{mn} \cos m\phi \cos n\psi + g_{mn} \cos m\phi \sin n\psi \nonumber\\ 
& + & ~~~~~~~~~~~~h_{mn} \sin m\phi \cos n\psi + i_{mn} \sin m\phi \sin n\psi)~.
\label{new_phipsi}
\end{eqnarray}
Here, $m$ and $n$ are the numbers of waves, $a$, $b_{m}$, $c_{m}$, $d_{n}$, $e_{n}$, $f_{mn}$, 
$g_{mn}$, $h_{mn}$, and $i_{mn}$ are the Fourier coefficients.
This equation includes cross terms in $\phi$ and $\psi$, while the original term 
in Eq.~(\ref{ene_torsion2}) has no mixing of $\phi$ and $\psi$.
Therefore, our new torsion-energy term can represent more complex energy surface 
than the conventional one.
The Fourier coefficients, by definition, are given by
\begin{equation}
c = \frac{1}{\alpha} \int_{-\pi}^{\pi} d\phi \int_{-\pi}^{\pi} d\psi~ 
{\cal E}(\phi,\psi) x(\phi,\psi) 
= \left( \frac{\pi}{180} \right)^2 \frac{1}{\alpha} \int_{-180}^{180} d\tilde{\phi} 
\int_{-180}^{180} d\tilde{\psi}~
{\cal E}\left( \frac{\pi}{180} \tilde{\phi},\frac{\pi}{180} \tilde{\psi} \right) 
x\left( \frac{\pi}{180} \tilde{\phi},\frac{\pi}{180} \tilde{\psi} \right)~,
\label{coe_eqn}
\end{equation}
where $\alpha$ are the normalization constants and $x(\phi, \psi)$ are the basis functions for 
the Fourier series.
Table~\ref{coe_table} summarizes these coefficients and functions.
Here, $\phi$ and $\psi$ are given in radians, and $\tilde{\phi}$ and $\tilde{\psi}$ are in degrees 
($\phi = \frac{\pi}{180} \tilde{\phi}$, $\psi = \frac{\pi}{180} \tilde{\psi}$). 
Hereafter, angular quantities without tilde and with tilde are in radians and in degrees,
respectively.

Finally, ${\cal E}(\phi,\psi)$ in Eq.~(\ref{new_phipsi}) and 
$E_{\rm rest}$ in Eq.~(\ref{ene_torsionb}) 
define our torsion-energy term in Eq.~(\ref{ene_conf}) (instead of Eq.~(\ref{ene_torsion})):
\begin{equation}
E_{\rm torsion} = {\cal E}(\phi,\psi) + E_{\rm rest}~.
\label{new_torsion}
\end{equation}

The double Fourier series in Eq.~(\ref{new_phipsi}) is particularly useful,
because it describes the backbone-torsion-energy surface in the Ramachandran space.
The Fourier series can express the torsion-energy surface ${\cal E}(\phi,\psi)$ 
that was obtained by any method including quantum chemistry calculations.
\cite{opls2,Carlos,Duan,IWA,MFB,Kamiya} 

Moreover, one can refine the existing backbone-torsion-energy term and 
control the secondary-structure-forming tendencies of the force fields.
For example, $\alpha$-helix is obtained 
for $(\tilde{\phi},\tilde{\psi}) \approx (-57^{\circ},-47^{\circ})$, 
$3_{10}$-helix
for $(\tilde{\phi},\tilde{\psi}) \approx (-49^{\circ},-26^{\circ})$, 
$\pi$-helix
for $(\tilde{\phi},\tilde{\psi}) \approx (-57^{\circ},-70^{\circ})$, 
parallel $\beta$-sheet 
for $(\tilde{\phi},\tilde{\psi}) \approx (-119^{\circ},113^{\circ})$, 
antiparallel $\beta$-sheet 
for $(\tilde{\phi},\tilde{\psi}) \approx (-139^{\circ},135^{\circ})$, 
and so on. \cite{Rama_Sasi}
Hence, if the existing force field gives, say, too little $\alpha$-helix-forming
tendency compared to experimental results, one can lower the backbone-torsion-energy 
surface near $(\tilde{\phi},\tilde{\psi}) = (-57^{\circ},-47^{\circ})$ 
in order to enhance
$\alpha$-helix formations.  

We can thus write 
\begin{equation}
{\cal E}(\phi,\psi) = E(\phi,\psi) - f(\phi,\psi)~,
\label{mod_surface}
\end{equation}
where
$E(\phi,\psi)$ is the existing backbone-torsion-energy term that we want to
refine and $f(\phi,\psi)$ is a function 
that has peaks around the corresponding regions where specific secondary structures 
are to be enhanced.
There are many possible choices for $f(\phi,\psi)$.
For instance, one can use the following function
when one wants to lower the torsion-energy surface
in a single region near $(\phi,\psi) = (\phi_0,\psi_0)$:
\begin{equation}
f(\phi,\psi) = 
\left \{ \begin{array}{ll} 
\displaystyle{A \exp \left( \frac{B}{(\phi - \phi_0)^2 + (\psi - \psi_0)^2 - {r_0}^2} \right) ~,}
& ~{\rm for}~~(\phi - \phi_0)^2 + (\psi - \psi_0)^2 < {r_0}^2~,\\[0.2cm]
                     0~, & ~{\rm otherwise}~,
                     \end{array} \right.
\label{mod_surface2}
\end{equation}
where $A$, $B$, and $r_0$ are constants that we adjust for refinement.
In this case, the energy surface is lowered by $f(\phi,\psi)$ in a circular region 
of radius $r_0$, which is centered at $(\phi,\psi) = (\phi_0,\psi_0)$.
Note that we should also impose periodic boundary conditions on $f(\phi,\psi)$.
   
We then express ${\cal E}(\phi,\psi)$ in Eq.~(\ref{mod_surface}) in terms of 
the double Fourier series in Eq.~(\ref{new_phipsi}),  
where the Fourier coefficients are obtained from Eq.~(\ref{coe_eqn}).
Hence, we can fine-tune the backbone-torsion-energy term by the
above procedure so that it yields correct secondary-structure-forming tendencies.

\section{Results and Discussion}
\label{results}

We now present the results of the applications of our backbone torsion energy
in Eq.~(\ref{new_phipsi}). 
In this section, we consider the following truncated Fourier series:
\begin{eqnarray}
{\cal E}(\phi,\psi) = a & + & b_{1} \cos \phi + c_{1} \sin \phi + b_{2} \cos 2\phi + c_{2} \sin 2\phi \nonumber\\ 
& + & d_{1} \cos \psi + e_{1} \sin \psi + d_{2} \cos 2\psi + e_{2} \sin 2\psi \nonumber\\ 
& + & f_{11} \cos \phi \cos \psi + g_{11} \cos \phi \sin \psi \nonumber\\ 
& + & h_{11} \sin \phi \cos \psi + i_{11} \sin \phi \sin \psi~.
\label{use_torsion}
\end{eqnarray}
This function has 13 Fourier-coefficient parameters.
We will see below that this number of Fourier terms is sufficient for
most of our purposes, but that for some cases more number of Fourier terms are preferred.

We first check how well the truncated Fourier series in Eq.~(\ref{use_torsion})
can reproduce the original AMBER parm94 and AMBER parm96 backbone-torsion-energy terms 
in Eqs.~(\ref{ene_torsion_parm94}) and (\ref{ene_torsion_parm96}).
Because these functions are already the sum of one-dimensional Fourier series
and subsets of the double Fourier series 
in Eq.~(\ref{new_phipsi}), the Fourier coefficients in Eq.~(\ref{coe_eqn}) can be
analytically calculated and agree with those 
in Eqs.~(\ref{ene_torsion_parm94}) and (\ref{ene_torsion_parm96})
except for the last one (that for $\cos 4 \psi$) 
in Eq.~(\ref{ene_torsion_parm94}).  This term is missing in Eq.~(\ref{use_torsion}).
These cases thus give us good test of numerical integrations in 
Eq.~(\ref{coe_eqn}).  The numerical integrations were evaluated as follows.
We divided the Ramachandran space 
($-180^{\circ} < \tilde{\phi} < 180^{\circ}$, 
$-180^{\circ} < \tilde{\psi} < 180^{\circ}$) 
into unit square cells of side length $\tilde{\epsilon}$ (in degrees).
Hence, there are $(360/\tilde{\epsilon})^2$ unit cells altogether.
The double integral on the right-hand side of Eq.~(\ref{coe_eqn}) was approximated by the sum of
$\left[{\cal E}\left( \frac{\pi}{180} \tilde{\phi},\frac{\pi}{180} \tilde{\psi} \right) 
x\left( \frac{\pi}{180} \tilde{\phi},\frac{\pi}{180} \tilde{\psi} \right)\right]
\times \left(\tilde{\epsilon}\right)^2$, 
where each 
${\cal E}\left( \frac{\pi}{180} \tilde{\phi},\frac{\pi}{180} \tilde{\psi} \right) 
x\left( \frac{\pi}{180} \tilde{\phi},\frac{\pi}{180} \tilde{\psi} \right)$ was
evaluated at one of the four corners of each unit cell.
We tried two values of $\tilde{\epsilon}$ ($1^{\circ}$ and $10^{\circ}$).
Both cases gave exact agreement of Fourier coefficients
with the results of the analytical integrations
for at least six digits (see Tables~\ref{coe_table2} and \ref{coe_table3} below).

In Fig.~\ref{fig_ene_surface_org}
we compare the backbone-torsion-energy surfaces of the original AMBER parm94 
and AMBER parm96 with those of
the corresponding double Fourier series in Eq.~(\ref{use_torsion}).
Hereafter, the primed labels for figures such as (a') indicate that
the results are those of the double Fourier series.
As can be seen from Figs.~\ref{fig_ene_surface_org}(b) and \ref{fig_ene_surface_org}(b'), 
the backbone-torsion-energy surfaces are in complete agreement for
AMBER parm96, whereas we see a little difference for AMBER parm94 between
Figs.~\ref{fig_ene_surface_org}(a) and \ref{fig_ene_surface_org}(a'). 
As discussed above, this slight difference for AMBER parm94 reflects the
fact that the $\cos 4 \psi$ term in Eq.~(\ref{ene_torsion_parm94})
is missing in the truncated double Fourier series in 
Eq.~(\ref{use_torsion}).

We now consider the double Fourier series of non-trigonometric functions.
The functions are those in 
Eqs.~(\ref{mod_surface}) and (\ref{mod_surface2}).
We try to fine-tune the original AMBER parm94 and AMBER parm96 force fields by
subtracting $f(\phi,\psi)$ in Eq.~(\ref{mod_surface2})
from the original functions.  The criterion for fine-tuning is, for instance,
whether the refined force fields yield better agreement of the 
secondary-structure-forming tendencies with experimental implications
than the original ones.
For this we need good experimental data.
Because the purpose of the present article is to test whether or not we can control
the secondary-structure-forming tendencies, we simply consider 
extreme cases where we try to modify the existing force fields so that
desired secondary structures may be obtained regardless of the
tendencies of the original force fields.  Note that the original AMBER parm94 and
AMBER parm96 favor $\alpha$-helix and $\beta$-sheet, respectively. \cite{YSO1,YSO2,SO1,SO2,SO3}

The function $f(\phi,\psi)$ in Eq.~(\ref{mod_surface2})
reduces the value of $E(\phi,\psi)$ in a circle of 
radius $r_0$ with the center located at $(\phi_0,\psi_0)$. 
We used $\tilde{r}_0 = 100^\circ$ and $\tilde{B}=5,000$ (degrees)$^2$. 
The coefficient $A$ is calculated by Eq.~(\ref{mod_surface2}) from the other parameters 
$f(\tilde{\phi}_0,\tilde{\psi}_0)$, $\tilde{r}_0$, and $\tilde{B}$. 
Namely, we have
\begin{equation}
A = f(\tilde{\phi}_0,\tilde{\psi}_0) \exp \left( \frac{\tilde{B}}{\tilde{r}_0^2} \right) ~.
\label{mod_surface3}
\end{equation}

We used
$(\tilde{\phi}_0,\tilde{\psi}_0) = (-57^{\circ},-47^{\circ})$ 
and 
$(\tilde{\phi}_0,\tilde{\psi}_0) = (-130^{\circ},125^{\circ})$ 
in order to enhance
$\alpha$-helix-forming tendency and $\beta$-sheet-forming tendency,
respectively.
The central values $f(\tilde{\phi}_0,\tilde{\psi}_0)$ that we used were
3.0 kcal/mol and 6.0 kcal/mol for enhancing $\alpha$-helix and
$\beta$-sheet, respectively, in the case of AMBER parm94.
They were both 3.0 kcal/mol in the case of AMBER parm96.
We remark that the large value of $f(\tilde{\phi}_0,\tilde{\psi}_0)$, 
6.0 kcal/mol, that was necessary to enhance $\beta$-sheet in the case of AMBER parm94
reflects the fact that the original force field favors $\alpha$-helix so much.

In Fig.~\ref{fig_ene_surface_mod} 
we show the backbone-torsion-energy surfaces modified according to
Eq.~(\ref{mod_surface}).  
We see that Eq.~(\ref{mod_surface}) reduces the torsion energy 
in the circular regions that correspond to the $\alpha$-helix region (a1 and b1)
and the $\beta$-sheet region (a2 and b2).
Hence, there are four cases: 
$\alpha$-helix is enhanced
from the original AMBER parm94 (a1) and AMBER parm96 (b1), 
and $\beta$-sheet is enhanced
from the original AMBER parm94 (a2) and AMBER parm96 (b2). 

These modified backbone-torsion-energy functions were expanded by the
truncated double Fourier series in Eq.~(\ref{use_torsion}) by evaluating the
corresponding Fourier coefficients from Eq.~(\ref{coe_eqn}).
For the numerical integration we again tried two values of 
the bin size $\tilde{\epsilon}$ ($1^{\circ}$ and $10^{\circ}$).
The obtained Fourier coefficients are summarized in
Table~\ref{coe_table2} in the case of AMBER parm94 and 
Table~\ref{coe_table3} in the case of AMBER parm96.
For comparisons, the Fourier coefficients of the original AMBER force fields
(before modifications) are also listed.
We see that the two choices of the bin size 
$\tilde{\epsilon}$ gave essentially the same results (up to
at least 3 digits).

In Fig.~\ref{fig_ene_surface_mod2} 
we show the backbone-torsion-energy surfaces represented by the
truncated double Fourier series.
Comparing these with the original ones in
Fig.~\ref{fig_ene_surface_mod}, we find that the overall features of the
energy surfaces are well reproduced by the Fourier series.
If more accuracy is desired, we can simply increase the number of 
Fourier terms in the expansion.
As we see below, the present accuracy of the Fourier series was sufficient for
the purpose of controlling the secondary-structure-forming tendencies
towards $\alpha$-helix or $\beta$-sheet. 

We examined the effects of the above modifications of the backbone-torsion-energy terms
in AMBER parm94 and AMBER parm96 (towards specific secondary structures) by performing
the folding simulations of two peptides, C-peptide of ribonuclease A and the 
C-terminal fragment of the B1 domain of streptococcal protein G, which is sometimes referred to as 
G-peptide \cite{gpep3}.
The C-peptide has 13 residues and its amino-acid sequence is
Lys-Glu-Thr-Ala-Ala-Ala-Lys-Phe-Glu-Arg-Gln-His-Met.
This peptide has been extensively studied by experiments and is known
to form an $\alpha$-helix structure \cite{buzz1,buzz2}.
Because the charges at peptide termini are known to affect
helix stability \cite{buzz1,buzz2}, we blocked the termini by
a neutral COCH$_3$- group and a neutral -NH$_2$ group.
The G-peptide has 16 residues and its amino-acid sequence is
Gly-Glu-Trp-Thr-Tyr-Asp-Asp-Ala-Thr-Lys-Thr-Phe-Thr-Val-Thr-Glu.
The termini were kept as the usual zwitter ionic states, following the
experimental conditions \cite{gpep3,gpep1,gpep2}.
This peptide is known to form a $\beta$-hairpin structure by experiments
\cite{gpep3,gpep1,gpep2}.

Simulated annealing \cite{SA} MD simulations were performed for both
peptides from fully extended initial conformations, where the four versions
of the truncated
double Fourier series (which were described in 
Tables~\ref{coe_table2} and \ref{coe_table3} and in
Fig.~\ref{fig_ene_surface_mod2}) 
were used for the backbone-torsion-energy terms
of AMBER parm94 and AMBER parm96 force fields.
For comparisons, the simulations 
with the original AMBER parm94 and parm96 force fields
were also performed. 
The unit time step was set to 1.0 fs.
Each simulation was carried out for 1 ns (hence, it consisted of
1,000,000 MD steps).
The temperature during MD simulations was controlled by 
Berendsen's method \cite{berendsen}.
For each run the temperature was decreased exponentially
from 2,000 K to 250 K.
As for solvent effects, we used the GB/SA model \cite{gb1,gb2}.
We modified and used the program package
TINKER version 4.1 \cite{tinker} for all the simulations.
For both peptides, these folding simulations were repeated 60 times 
with different sets of randomly generated initial velocities.

In Table~\ref{table-c-pep}, the numbers of obtained conformations (final conformations)
with $\alpha$-helix structures and $\beta$-hairpin 
structures are listed for the case of C-peptide, which is known to form
$\alpha$-helix structures by experiments. 
We used DSSP \cite{DSSP} for the criterions of secondary-structure formations.
We see that all 60 conformations obtained from the simulations using the original AMBER parm94 are 
$\alpha$-helix structures and that for the original AMBER parm96, 14 out of 60 conformations are 
$\alpha$-helix structures and 16 out of 60 conformations are $\beta$-hairpin structures.
The results confirm that the original AMBER parm94 strongly
favors $\alpha$-helix structures and that the original 
AMBER parm96 slightly favors $\beta$-hairpin structures (because this peptide
should form $\alpha$-helices). \cite{YSO1,SO1}
However, for both AMBER parm94 and AMBER parm96 
modified to enhance $\alpha$-helix-forming tendency, 
almost all (60 and 59) conformations 
are $\alpha$-helix structures
and there are no conformations with $\beta$-hairpin structures. 
For AMBER parm94 and AMBER parm96 modified to enhance $\beta$-sheet-forming tendency, 
on the other hand, about half the conformations exhibit
$\beta$-hairpin structures and no $\alpha$-helix structures are found.
   
\begin{figure}
\begin{center}
\includegraphics[width=10cm,keepaspectratio]{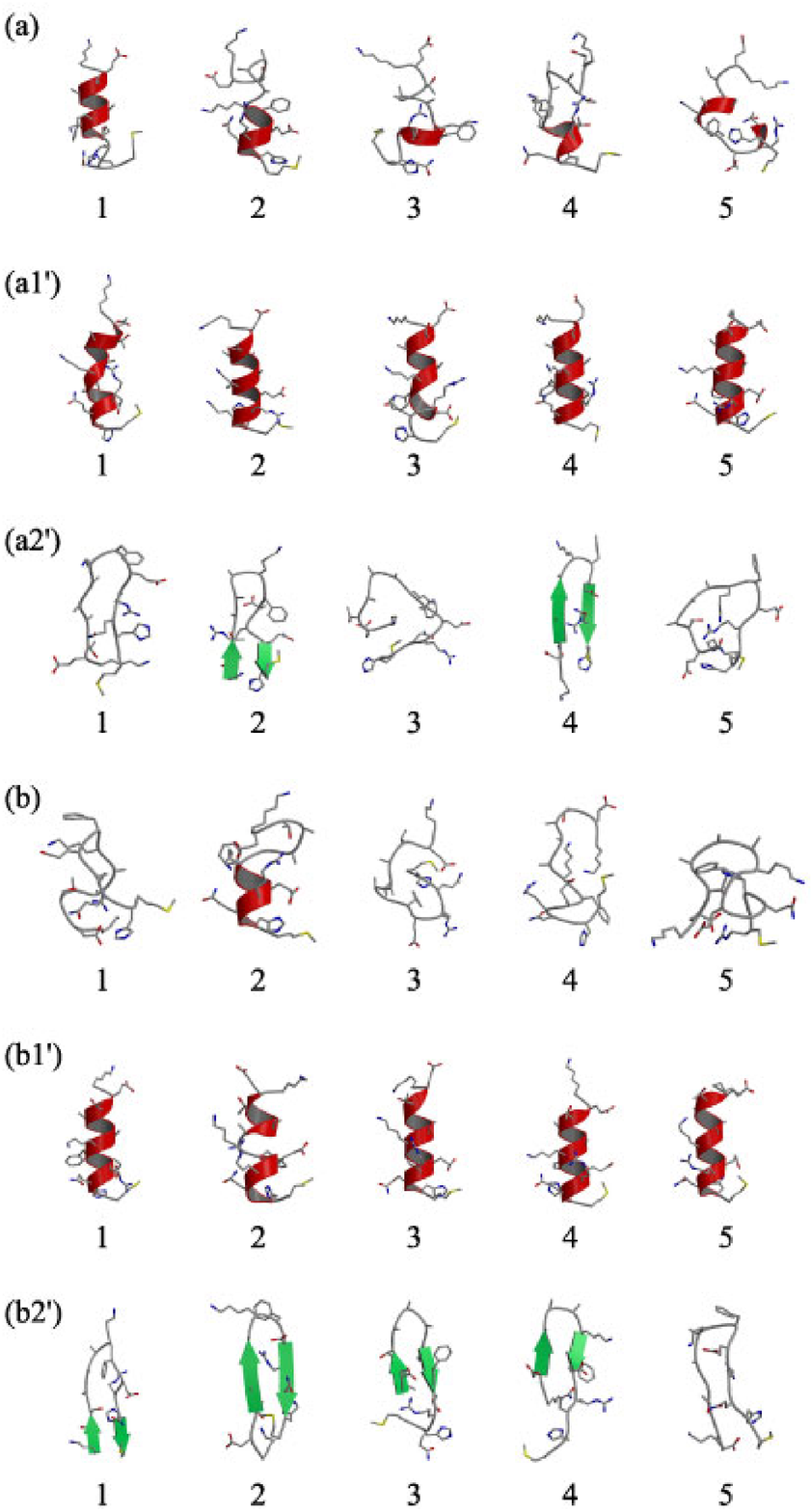}
\caption{Five lowest-energy final conformations of C-peptide obtained from six sets of
60 simulated annealing MD runs.
(a) and (b) are the results of the original AMBER parm94 and the original AMBER parm96
force fields, respectively.
(a1') and (a2') are those of the truncated double Fourier series of
AMBER parm94 force fields that were modified to
enhance $\alpha$-helix structures and $\beta$-sheet structures, respectively.
(b1') and (b2') are those of the truncated double Fourier series of
AMBER parm96 force fields that were modified to
enhance $\alpha$-helix structures and $\beta$-sheet structures, respectively.
The conformations are ordered in the increasing order of energy for each case.  
The figures were created with Molscript\cite{molscript} and Raster3D\cite{raster3d}.}
\label{fig_94_96_cpep}
\end{center}
\end{figure}

In Fig.~\ref{fig_94_96_cpep} we show five (out of 60) lowest-energy final conformations of 
C-peptide obtained by the simulated annealing MD simulations for the six cases.
According to Table~\ref{table-c-pep}, there are almost 100 \% $\alpha$-helix formations 
in three cases, namely, the original AMBER parm94, the AMBER parm94 modified towards
$\alpha$-helix, and the AMBER parm96 modified towards $\alpha$-helix.
Here in Fig.~\ref{fig_94_96_cpep} we see the differences in the three cases:
the modified AMBER parm94 and parm96 favor $\alpha$-helix structures more than the
original AMBER parm94 in the sense that the obtained helices are more extended 
(and almost entirely helical) in the former cases.
Moreover, we see clear $\beta$-hairpin formations with extended $\beta$ strands
in the cases of the AMBER parm94 and AMBER parm96 modified towards $\beta$-sheet.

\begin{figure}
\begin{center}
\includegraphics[width=10cm,keepaspectratio]{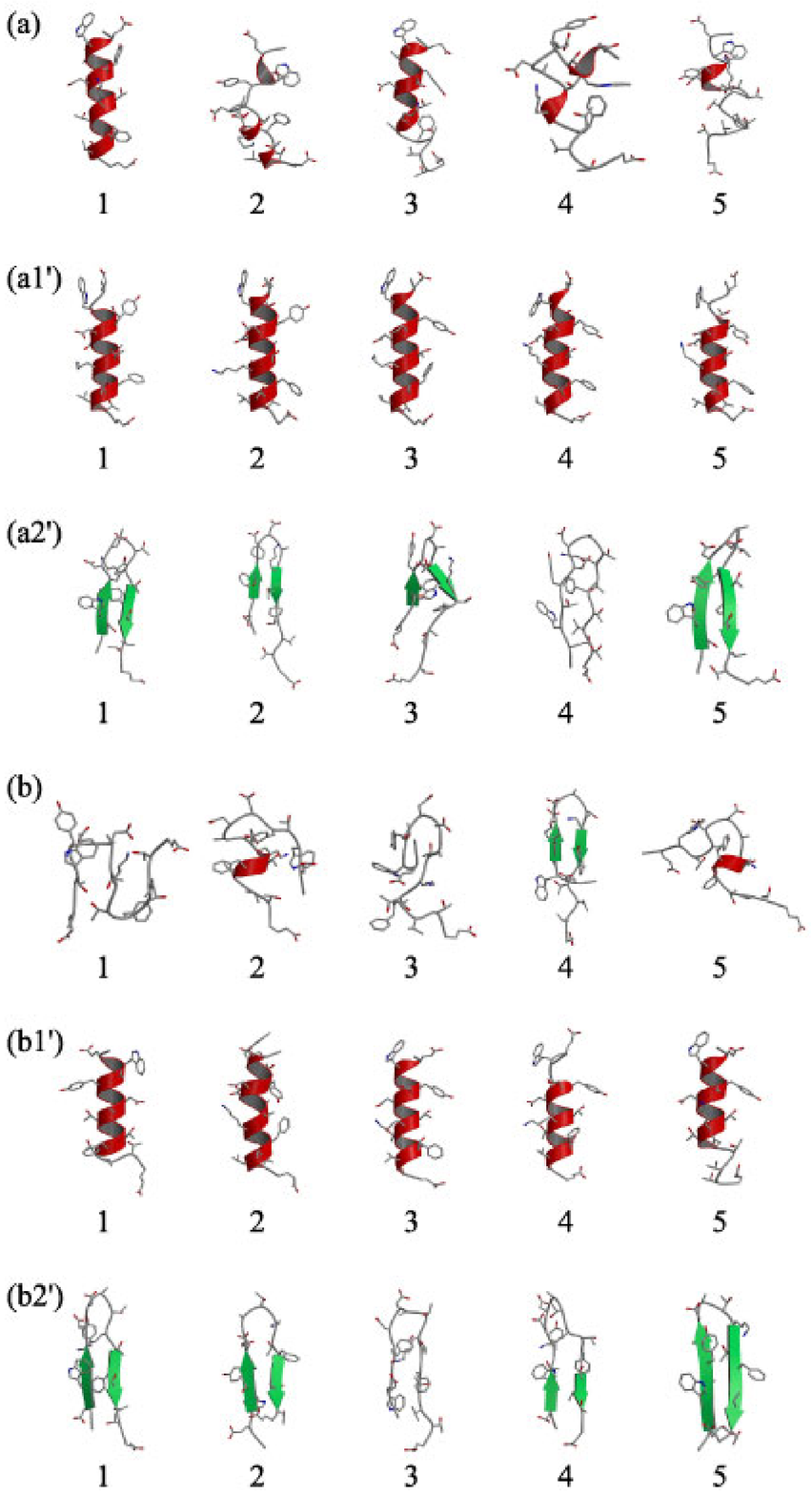}
\caption{Five lowest-energy final conformations of G-peptide obtained from six sets of
60 simulated annealing MD runs.
(a) and (b) are the results of the original AMBER parm94 and the original AMBER parm96
force fields, respectively.
(a1') and (a2') are those of the truncated double Fourier series of
AMBER parm94 force fields that were modified to
enhance $\alpha$-helix structures and $\beta$-sheet structures, respectively.
(b1') and (b2') are those of the truncated double Fourier series of
AMBER parm96 force fields that were modified to
enhance $\alpha$-helix structures and $\beta$-sheet structures, respectively.
The conformations are ordered in the increasing order of energy for each case.  
The figures were created with Molscript\cite{molscript} and Raster3D\cite{raster3d}.}
\label{fig_94_96_gpep}
\end{center}
\end{figure}

In Table~\ref{table-g-pep}, the numbers of obtained conformations (final conformations)
with $\alpha$-helix structures and $\beta$-hairpin 
structures are listed for the case of G-peptide, which is known to form
$\beta$-hairpin structures by experiments. 
The results are similar to those in Table~\ref{table-c-pep}.
We see that all 60 conformations are $\alpha$-helix structures
in the cases of the original 
AMBER parm94 and the AMBER parm94 and AMBER parm96 that were modified to enhance
$\alpha$-helix structures. 
For AMBER parm94 and AMBER parm96 that were modified to enhance $\beta$-sheet-forming tendency, 
the conformations exhibit $\beta$-hairpin structures and no $\alpha$-helix structures are found.
One difference in the results of the two Tables is that
the original AMBER parm96 clearly favors $\beta$-hairpin structures in the case of
G-peptide, while in the case of C-peptide about the same tendency is observed
for $\alpha$-helix and $\beta$-hairpin with this original force field.
Our modifications of the force fields resulted in the expected changes in
the secondary-structure formations.  This is again more clearly shown in
Fig.~\ref{fig_94_96_gpep} where five lowest-energy final conformations of G-peptide 
are displayed for the six cases.  Overall features are the same as in 
Fig.~\ref{fig_94_96_cpep}.

Therefore, regardless of the secondary-structure-forming tendencies of 
the original force fields, our modifications of the backbone-torsion-energy term 
succeeded in enhancing the desired secondary structures.

\section{Conclusions}
\label{conclusions}
In this article, we proposed to
use a double Fourier series in two backbone dihedral angles, 
$\phi$ and $\psi$, for the backbone-torsion-energy term in
protein force fields.
This is a natural generalization of the conventional torsion-energy terms.
It is particularly useful in controlling secondary-structure-forming
tendencies, because any function in the Ramachandran space can be
expanded by this double Fourier series.
We can easily modify the existing force fileds so that
specified secondary-struture-forming tendencies are enhanced, by
lowering the backbone-torsion-energy surface in the corresponding region 
of the Ramachandran space.
We demonstrated this taking the examples of AMBER parm94 and AMBER parm96 force
fields.

Besides the above ``manual'' adjustment of the force fields, we can also apply
our force-field refinement method \cite{SO1,SO2,SO3} to this
double Fourier series.  Namely, we can determine the values of the
Fourier coefficients so that the forces acting on the atoms with the coodinates of
the PDB database become minimal.  Work is underway in this direction.

\newpage
\noindent
{\bf Acknowledgements}: \\
The computations were performed on SuperNOVA of Miki Laboratory at Doshisha University
and the computers at the Research Center for Computational Science, Institute for Molecular Science.
This work was supported, in part, by 
the Grants-in-Aid for the Academic Frontier Project, ``Intelligent Information Science'', 
for the Nano-Bio-IT Education Program, 
for the NAREGI Nanoscience Project, and for Scientific Research in Priority Areas, ``Water and Biomolecules'', from the 
Ministry of Education, Culture, Sports, Science
and Technology, Japan.



\newpage

\begin{table}
\caption{Torsion-energy parameters for the backbone dihedral angles $\phi$ and $\psi$ 
for AMBER parm94 and AMBER parm96 in Eq.~(\ref{ene_torsion2}). }
\label{table-org-torsion}
\vspace{0.3cm}
\begin{tabular}{lcccccc} \hline
  &   & $\phi$ &            &     & $\psi$ &            \\ \hline
  & $m$ & $\displaystyle{\frac{V_m}{2}}$ (kcal/mol) & $\gamma_m$ (radians)~~~~~ & $n$ & $\displaystyle{\frac{V_n}{2}}$ (kcal/mol) & $\gamma_n$ (radians)\\ \hline
parm94~~~~~ &  2  & 0.2 & $\pi$ & 1  & 0.75 & $\pi$  \\
       &     &       &            &  2  & 1.35 & $\pi$ \\
       &     &       &            &  4  & 0.4 & $\pi$      \\ \hline
parm96~~~~~ &  1  & 0.85 &   0   &  1  & 0.85 &   0   \\ 
       &  2  & 0.3 & $\pi$  &  2  & 0.3 & $\pi$   \\ \hline
\end{tabular}
\end{table}

\newpage

\begin{table}
\caption{Fourier coefficients $c$, normalization constants $\alpha$, and the basis functions 
$x(\phi, \psi)$ for the double Fourier series of the backbone torsion energy ${\cal E}(\phi,\psi)$ 
in Eqs.~(\ref{new_phipsi}) and (\ref{coe_eqn}).}
\label{coe_table}
\vspace{0.3cm}
\begin{tabular}{lrr} \hline
$c$      &  ~~$\alpha$  &  ~~~~~~~~~~~~~$x(\phi, \psi)$   \\ \hline
$a$      & $4\pi^2$   &  1                 \\
$b_{m}$  & $2\pi^2$   &  $\cos m\phi$         \\
$c_{m}$  & $2\pi^2$   &  $\sin m\phi$         \\
$d_{n}$  & $2\pi^2$   &  $\cos n\psi$         \\
$e_{n}$  & $2\pi^2$   &  $\sin n\psi$         \\
$f_{mn}$ & $\pi^2$    &  $\cos m\phi \cos n\psi$ \\
$g_{mn}$ & $\pi^2$    &  $\cos m\phi \sin n\psi$ \\
$h_{mn}$ & $\pi^2$    &  $\sin m\phi \cos n\psi$ \\
$i_{mn}$ & $\pi^2$    &  $\sin m\phi \sin n\psi$ \\ \hline
\end{tabular}
\end{table}

\newpage

\begin{table}
\caption{Fourier coefficients in Eq.~(\ref{use_torsion}) obtained from
the numerical evaluations of the integrals in Eq.~(\ref{coe_eqn}).
``org94'' stands for the original AMBER parm94 force field.
``mod94($\alpha$)'' and ``mod94($\beta$)'' stand for AMBER parm94 force fields 
that were modified to
enhance $\alpha$-helix structures and $\beta$-sheet structures, respectively,
by Eqs.~(\ref{mod_surface}) and (\ref{mod_surface2}).
The bin size $\tilde{\epsilon}$ is the length of the sides of each unit
square cell for the numerical integration in Eq.~(\ref{coe_eqn}).}
\label{coe_table2}
\vspace{0.3cm}
\begin{tabular}{lrrrrrr} \hline
bin size $\tilde{\epsilon}$   &             & $1^\circ$       &                &             & $10^\circ$      &                 \\ \hline
coefficient & org94    &  mod94($\alpha$) &  mod94($\beta$) &  org94   &  mod94($\alpha$) &  mod94($\beta$)  \\ \hline
$a$         &  2.700000   &  2.308359       &  2.308359      &  2.700000   &  2.308370       &  2.308371       \\
$b_1$       &  0.000000   &$-0.330937$      &  0.390575      &  0.000000   &$-0.331053$      &  0.390521       \\
$b_2$       &$-0.200000$  &$-0.101549$      &$-0.157968$     &$-0.200000$  &$-0.101513$      &$-0.157985$      \\
$c_1$       &  0.000000   &  0.509599       &  0.465469      &  0.000000   &  0.509517       &  0.465404       \\
$c_2$       &  0.000000   &  0.221123       &$-0.238372$     &  0.000000   &  0.221100       &$-0.238279$      \\
$d_1$       &$-0.750000$  &$-1.164401$      &$-0.401480$     &$-0.750000$  &$-1.164500$      &$-0.401437$      \\
$d_2$       &$-1.350000$  &$-1.333115$      &$-1.267214$     &$-1.350000$  &$-1.333073$      &$-1.267170$      \\
$e_1$       &  0.000000   &  0.444390       &$-0.497739$     &  0.000000   &  0.444289       &$-0.497800$      \\
$e_2$       &  0.000000   &  0.241460       &  0.227452      &  0.000000   &  0.241451       &  0.227573       \\ 
$f_{11}$    &  0.000000   &$-0.342789$      &$-0.340247$     &  0.000000   &$-0.343087$      &$-0.340249$      \\ 
$g_{11}$    &  0.000000   &  0.367596       &  0.485922      &  0.000000   &  0.367697       &  0.485925       \\ 
$h_{11}$    &  0.000000   &  0.527849       &$-0.405490$     &  0.000000   &  0.527949       &$-0.405492$      \\ 
$i_{11}$    &  0.000000   &$-0.566049$      &  0.579100      &  0.000000   &$-0.565751$      &  0.579103       \\ \hline
\end{tabular}
\end{table}

\newpage

\begin{table}
\caption{Fourier coefficients in Eq.~(\ref{use_torsion}) obtained from
the numerical evaluations of the integrals in Eq.~(\ref{coe_eqn}).
``org96'' stands for the original AMBER parm96 force field.
``mod96($\alpha$)'' and ``mod96($\beta$)'' stand for AMBER parm96 force fields 
that were modified to
enhance $\alpha$-helix structures and $\beta$-sheet structures, respectively,
by Eqs.~(\ref{mod_surface}) and (\ref{mod_surface2}).
See also the caption of Table~\ref{coe_table2}.}
\label{coe_table3}
\vspace{0.3cm}
\begin{tabular}{lrrrrrr} \hline
bin size $\tilde{\epsilon}$   &             & $1^\circ$       &                &             & $10^\circ$      &                 \\ \hline
coefficient & org96    &  mod96($\alpha$) &  mod96($\beta$) &  org96   &  mod96($\alpha$) &  mod96($\beta$)  \\ \hline
$a$         &  2.300000   &  1.908359       &  1.908359      &  2.300000   &  1.908370       &  1.908371       \\
$b_1$       &  0.850000   &  0.519063       &  1.240575      &  0.850000   &  0.518947       &  1.240521       \\
$b_2$       &$-0.300000$  &$-0.201549$      &$-0.257968$     &$-0.300000$  &$-0.201513$      &$-0.257985$      \\
$c_1$       &  0.000000   &  0.509599       &  0.465469      &  0.000000   &  0.509517       &  0.465404       \\
$c_2$       &  0.000000   &  0.221123       &$-0.238372$     &  0.000000   &  0.221100       &$-0.238279$      \\
$d_1$       &  0.850000   &  0.435599       &  1.198520      &  0.850000   &  0.435500       &  1.198563       \\
$d_2$       &$-0.300000$  &$-0.283115$      &$-0.217214$     &$-0.300000$  &$-0.283073$      &$-0.217170$      \\
$e_1$       &  0.000000   &  0.444390       &$-0.497739$     &  0.000000   &  0.444289       &$-0.497800$      \\
$e_2$       &  0.000000   &  0.241460       &  0.227452      &  0.000000   &  0.241451       &  0.227573       \\ 
$f_{11}$    &  0.000000   &$-0.342789$      &$-0.340247$     &  0.000000   &$-0.343087$      &$-0.340249$      \\ 
$g_{11}$    &  0.000000   &  0.367596       &  0.485922      &  0.000000   &  0.367697       &  0.485925       \\ 
$h_{11}$    &  0.000000   &  0.527849       &$-0.405490$     &  0.000000   &  0.527949       &$-0.405492$      \\ 
$i_{11}$    &  0.000000   &$-0.566049$      &  0.579100      &  0.000000   &$-0.565751$      &  0.579103       \\ \hline
\end{tabular}
\end{table}

\newpage

\begin{table}
\caption{Number of final conformations with secondary structures obtained from the folding simulations 
of C-peptide. The total number of folding simulations in each case was 60.
See the captions of Tables~\ref{coe_table2} and \ref{coe_table3}.
``$\alpha$-helix'' stands for the number of conformations including the amino acids 
which were identified to be ``H''(= $\alpha$-helix) by DSSP.
``$\beta$-hairpin'' stands for the number of conformations including the amino acids 
which were identified to be ``B''(= residue in isolated $\beta$-bridge) 
or ``E''(= extended strand, participating in $\beta$-ladder) by DSSP.}
\label{table-c-pep}
\vspace{0.3cm}
\begin{tabular}{lrrrrrr} \hline
secondary structure & org94 & mod94($\alpha$) & mod94($\beta$) & ~~~~~org96 & mod96($\alpha$) & mod96($\beta$) \\ \hline
$\alpha$-helix  & 60/60  & 60/60 & 0/60 & 14/60 & 59/60 & 0/60 \\
$\beta$-hairpin & 0/60 & 0/60 & 29/60 & 16/60 & 0/60 & 34/60 \\ \hline
\end{tabular}
\end{table}

\newpage

\begin{table}
\caption{Number of final conformations with secondary structures obtained from the folding simulations 
of G-peptide.  See the caption of Table~\ref{table-c-pep}.}
\label{table-g-pep}
\vspace{0.3cm}
\begin{tabular}{lrrrrrr} \hline
secondary structure & org94 & mod94($\alpha$) & mod94($\beta$) & ~~~~~org96 & mod96($\alpha$) & mod96($\beta$) \\ \hline
$\alpha$-helix  & 60/60  & 60/60 & 0/60 & 12/60 & 60/60 & 0/60 \\
$\beta$-hairpin & 0/60 & 0/60 & 31/60 & 23/60 & 0/60 & 22/60 \\ \hline
\end{tabular}
\end{table}


\begin{figure}
\begin{center}
\includegraphics[width=11cm,keepaspectratio]{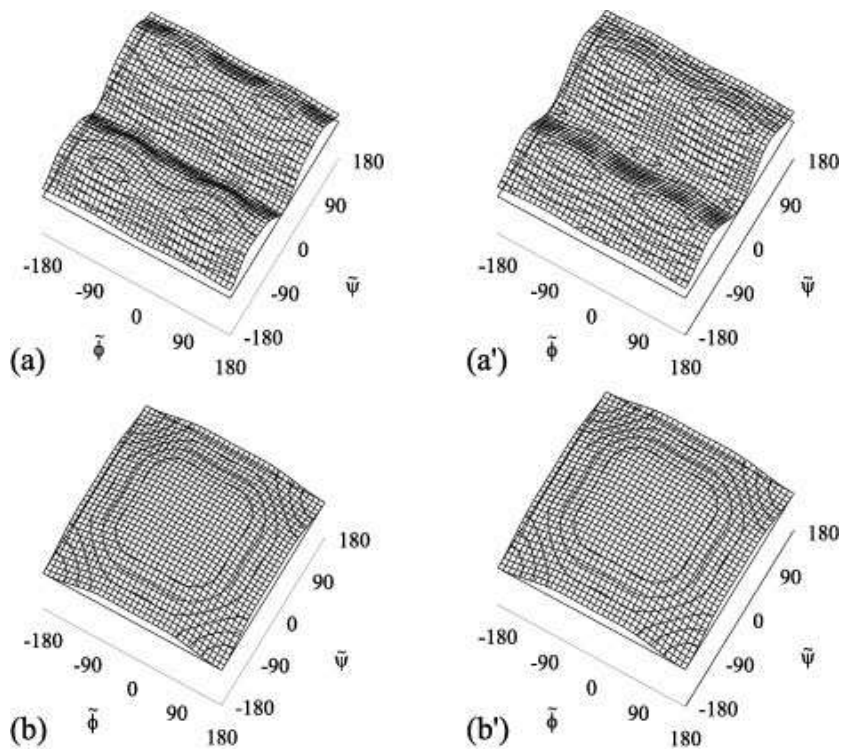}
\caption{Backbone-torsion-energy surfaces of AMBER force fields.  The backbone dihedral
angles $\tilde{\phi}$ and $\tilde{\psi}$ are in degrees. 
(a) and (b) are those of the original AMBER parm94 and the original AMBER parm96, respectively. 
(a') and (b') are those of (a) and (b), respectively, that are expressed by the truncated double 
Fourier series in Eq.~(\ref{use_torsion}).  The contour lines are drawn every 0.5 kcal/mol.}
\label{fig_ene_surface_org}
\end{center}
\end{figure}


\begin{figure}
\begin{center}
\includegraphics[width=11cm,keepaspectratio]{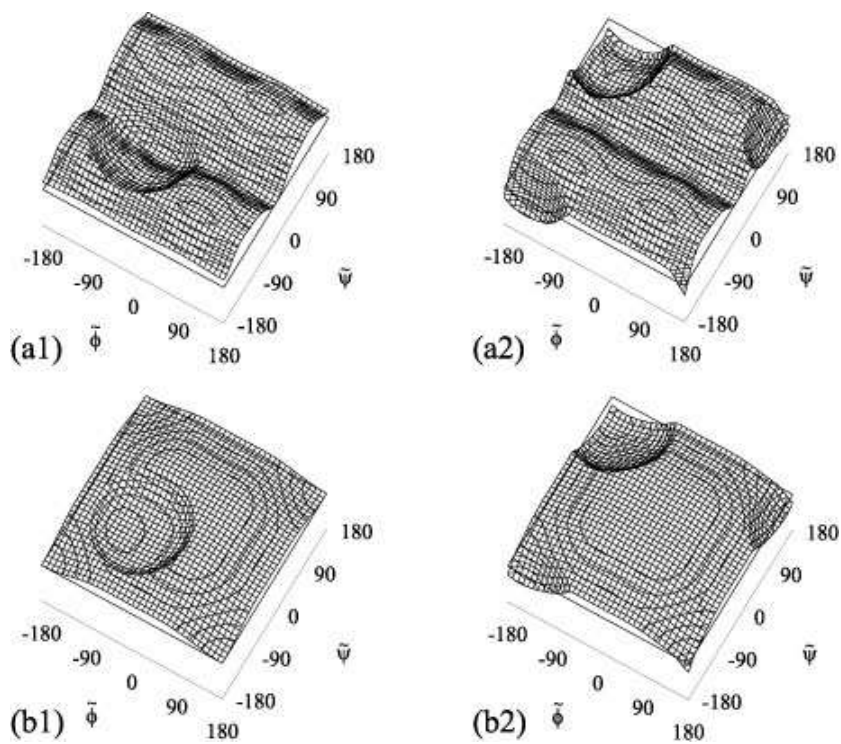}
\caption{Backbone-torsion-energy surfaces of AMBER force fields that were modified by
Eqs.~(\ref{mod_surface}) and (\ref{mod_surface2}).
(a1) and (a2) are those of AMBER parm94 force fields that were modified to
enhance $\alpha$-helix structures and $\beta$-sheet structures, respectively.
(b1) and (b2) are those of AMBER parm96 force fields that were modified to
enhance $\alpha$-helix structures and $\beta$-sheet structures, respectively.
See also the caption of Fig.~\ref{fig_ene_surface_org}.}
\label{fig_ene_surface_mod}
\end{center}
\end{figure}


\begin{figure}
\begin{center}
\includegraphics[width=11cm,keepaspectratio]{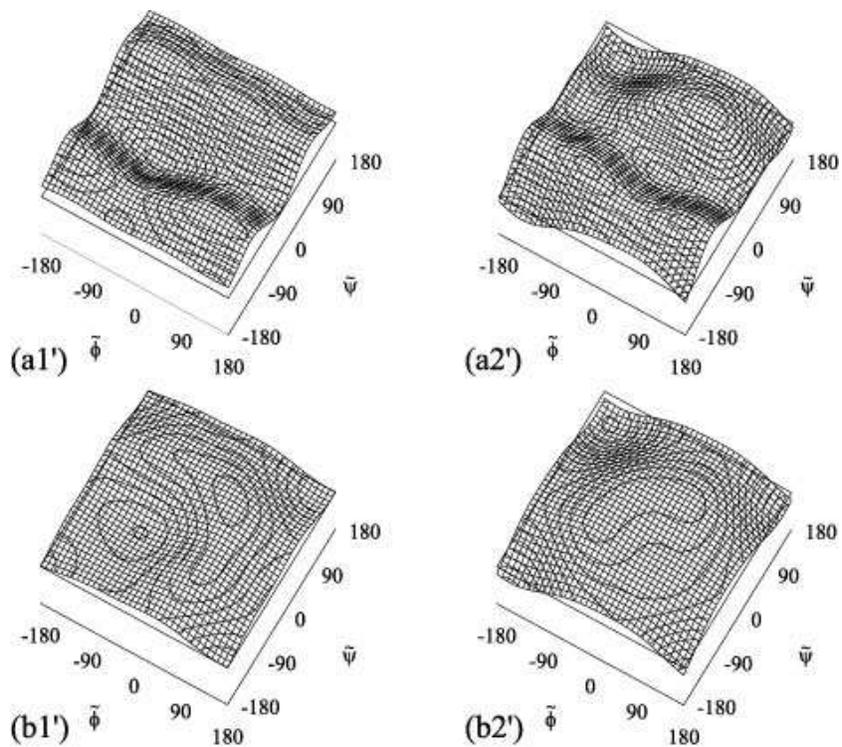}
\caption{Backbone-torsion-energy surfaces of AMBER force fields that were modified by
Eqs.~(\ref{mod_surface}) and (\ref{mod_surface2}) and expanded by the truncated
double Fourier series in Eq.~(\ref{use_torsion}).
(a1') and (a2') are those of AMBER parm94 force fields that were modified to
enhance $\alpha$-helix structures and $\beta$-sheet structures, respectively.
(b1') and (b2') are those of AMBER parm96 force fields that were modified to
enhance $\alpha$-helix structures and $\beta$-sheet structures, respectively.
See also the caption of Fig.~\ref{fig_ene_surface_org}.}
\label{fig_ene_surface_mod2}
\end{center}
\end{figure}

\newpage

\newpage

\end{document}